\documentclass[epjST]{svjour}
\usepackage{graphics,graphicx,amssymb,ulem}
\usepackage{hyperref}
\usepackage{bm}
\usepackage{slashed} 
\usepackage[utf8]{inputenc}
\usepackage{ amssymb } 
\usepackage{amsmath}
\usepackage{xcolor}
\newcommand{\Lagr}{\mathcal{L}}

\newcommand{\dd}{\textrm{d}}
\begin{document}
\title{Studying the parameters of the extended $\sigma$-$\omega$ model for neutron star matter}
\author{David~Alvarez-Castillo\inst{1,2}
\and
Alexander~Ayriyan\inst{3,4}
\and
Gergely~G\'abor~Barnaf\"{o}ldi\inst{5}
\and
Hovik~Grigorian\inst{3,4,6}
\and
P\'eter~P\'osfay\inst{5}
}                     
\offprints{}          
\institute{
 Henryk Niewodnicza\'nski Institute of Nuclear Physics, Cracow, Poland
 \and
Bogoliubov Laboratory of Theoretical Physics, JINR, Dubna, Russia
\and
Laboratory of Information Technologies, JINR Dubna, Dubna, Russia
\and 
Computational Physics and IT Division, A.I. Alikhanyan National Science Laboratory, Armenia
\and
Wigner Research Centre for Physics, Budapest, Hungary
\and
Department for Theoretical Physics, Yerevan State University, Yerevan, Armenia
}
\date{Received: date / Revised version: date}
%
\abstract{
In this work we study the parameters of the extended $\sigma$-$\omega$ model for neutron star matter by a Bayesian analysis of state-of-the-art multi-messenger astronomy observations, namely mass, radius and tidal deformabilities. We have considered three parameters of the model, the Landau mass $m_L$, the nuclear compressibility $K_0$, and the value of the symmetry energy $S_0$, all at saturation density $n_0$.
 As a result, we are able to estimate the values of the Landau mass of f $m_L = 739\pm17$ MeV, whereas the values of $K_0$ and $S_0$ fall within already known empirical values. Furthermore, for neutron stars we find the most probable value of 13 km $<R_{1.4}<$ 13.5 km and the upper mass limit of $M_{max} \approx 2.2$ M$_{\odot}$.
%
} 
\maketitle
\section{Introduction}
\label{intro}

During the recent years the uncertainty on the determination of the equation of state (EoS) of neutron stars has been reduced by both laboratory experiments and by multi-messenger astronomy observations. In the neutron star interior matter is as dense as several times saturation density $n_0$, the mean density of heavy atomic nuclei. Studying this broad range of densities represents a great challenge for both experiments and theories. 
The properties of nuclear matter around saturation density have been measured in different experiments like is the case of relativistic heavy ion collisions, giant dipole resonances of nuclei, parity violating weak neutral interaction scattering on heavy nuclei, among others~\cite{Tsang:2012se,Newton:2011aa}. Higher densities are hard to study in the laboratory particularly in the case of cold and neutron rich nuclear matter inside compact stars. On the contrary, atomic nuclei are mostly symmetric in their proton and neutron content. Observations of neutron stars may provide estimates of their masses and radii as well as other properties like their deformation, which is intimately related to gravitational wave emissions, see~\cite{Baiotti:2019sew} for an overview.
From the theoretical front, nuclear matter at high densities harbors the possibility that neutron star interiors bear exotic content beyond protons and neutrons, like hyperons, meson condensates or deconfined quark matter, resulting in different predictions for mass and radius relations. It is in this high density part where lies the discrepancy between the many EoS models, with many extrapolating from their low density results.  
In our study, we consider canonical neutron stars composed of protons and neutrons as well as leptons that account for charge neutrality. As a theoretical tool, the extended  $\sigma $-$\omega$ relativistic model~\cite{Posfay:2020xgp} is solved in its mean field version. We rely on the simplicity of this EoS model in order to obtain a robust framework where parameter changes can control the high density EoS extrapolation thus able to cover regions of interest in the mass-radius diagram for neutron stars.
 We choose three parameters to characterise the neutron star EoS which we estimate through a Bayesian analysis based on astronomical observations: the mass measurements of the PSR\,J0740+6620~\cite{Cromartie:2019kug} and PSR\,J0348+0432~\cite{Antoniadis:2013pzd}, the tidal deformability estimates from the gravitational wave event GW170817~\cite{TheLIGOScientific:2017qsa,Abbott:2018exr} and maximum mass boundary from its electromagnetic signal counterpart~\cite{Most:2018hfd}, as well as the X-ray measurement of the PSR J0030+0451 by the NICER detector~\cite{Miller:2019cac,Riley:2019yda}.

The EoS parameters we consider here are the compressibility of nuclear matter $K_0$, the nuclear symmetry $S_0$, and the Landau mass $m_L$, all evaluated at saturation density. These first two quantities have already been measured in the laboratory with resulting values of about $K_0 \approx 240 \pm 20$ MeV~\cite{Shlomo:2016ola,Piekarewicz:2009gb} and $S_0 = 31.7 \pm 3.2$ MeV~\cite{Li:2013ola,Oertel:2016bki}. We shall compare these empirical values with the ones resulting from our Bayesian analysis based on neutron star observations. Moreover, the Landau mass is an effective mass resulting from the interaction of nucleons in the dense medium. In a primer study~\cite{Alvarez-Castillo:2020aku} we have estimated its value to be of  $m_L = 750 \pm30$ MeV where we varied only one parameter while keeping the others fixed, with values of $n_0=0.156$ fm$^{-3}$, $K_0=240$ MeV, and $S_0=32.5$ MeV. Consequently, a generalization of that method to three parameters is presented in this work.
 
The compressibility of nuclear matter $K$ has a dominant effect on the stiffness of the EoS which is translated into the maximum neutron star mass $M_{max}$. While the symmetry energy $S$ strongly affects the size of neutron stars, it also fully determines the proton fraction in their interiors. Cooling of neutron stars may undergo the fast DUrca cooling~\cite{Lattimer:1991ib} if the proton fraction goes above some threshold value. From population observations, this fast cooling is not favoured~\cite{Popov:2004ey}. Thus, the link  between the symmetry energy and temperature observations of compact stars, see~\cite{Blaschke:2004vq,Grigorian:2016leu,Grigorian:2017xqd} and references therein. Most importantly, a correlation between the neutron star radius and the tidal deformability has been found in \cite{Horowitz:2019piw,Tong:2019juo}. 
 
Bayesian methods have been widely used for estimation of model parameters as well as for statistical inference. In the case of the neutron star EoS studies, we can briefly summarise a few works
before and after the multi-messenger era, i.e. before and after the GW170817 detection when many electromagnetic counterparts complemented the gravitational wave observation.  As for earlier works it is worth to mention the seminal work which performed an X-ray bursters Bayesian analysis~\cite{Steiner:2010fz,Raithel:2017ity}. Raithel et al.~\cite{Raithel:2017ity} introduced the popular multi-polytrope EoS parameterization (MPP)~\cite{Zdunik:2005kh,Read:2008iy,Alvarez-Castillo:2017qki} whereas~\cite{Alvarez-Castillo:2016oln} exploited realistic EoS instead. Moreover, Lackey and Wade~\cite{Lackey:2014fwa} considered future detections of gravitation radiation from binary neutron stars populations with a four parameter polytrope formulation EoS concluding that the high density EoS could be more accurately determined. With the advent of the multi-messenger astronomy era many other Bayesian studies appeared, leading to compatible results. For instance~\cite{Most:2018hfd}, presents an exhaustive study using the MPP approach to cover the phase space of EoS parameters by imposing constraints on the lower bound of the maximum neutron star mass and tidal deformability values. Later on, the mass-radius measurement of PSR J0030+0451 by NICER would help to provide tighter constraints on the resulting EoS regions, see~\cite{Ayriyan:2018blj,Miller:2019nzo,Raaijmakers:2018bln,Raaijmakers:2019dks,Traversi:2020aaa,Capano:2019eae,Lim:2019som,Essick:2019ldf,LIGOScientific:2019eut,Essick:2020flb} where different considerations and constraints, for example different Bayesian priors, have been taken into account. Most importantly, the EoS study parameters considered here have been studied in a few other Bayesian works that consider different EoS models, as is the case of Miller et al.~\cite{Miller:2019nzo} that use the MPP EoS parameterization to study the impact of laboratory measured the symmetry energy on the compact star EoS, below nuclear saturation densities.


The Bayesian methodology used in this work is a continuation of previous analyses~\cite{Alvarez-Castillo:2016oln,Ayriyan:2018blj} where the EoS modeled hybrid stars with hadron matter undergoing a phase transition into deconfined quark matter
described either by a Maxwell construction or a mixed phase.  A very important aspect shared in these Bayesian studies is the implementation of an unrestricted lower bound for the maximum neutron star mass, a quantity that keeps changing due to the observational updates, therefore influencing the posterior probabilities results~\cite{Alvarez-Castillo:2016oln}. 
The accompanying analysis~\cite{Blaschke:2020qqj} to our study here a) focuses on EoS models that show a third branch in the $M$-$R$ diagram~\cite{Benic:2014jia,Montana:2018bkb,Alvarez-Castillo:2018pve} and, b) implements the same observational constraints and Bayesian methodology used here. In addition, a more detailed overview of the state-of-the-art neutron star Bayesian studies can be found there. Alternatively, an EoS study using deep neural networks has been presented in~\cite{Fujimoto:2019hxv} bearing similar results to most Bayesian analyses.

This paper is organized as follows. In section~\ref{EoS} we introduce the extended $\sigma $-$\omega$ model for the neutron star EoS. In section~\ref{Sequences} we briefly describe the methods used to compute neutron star parameters followed by section~\ref{Bayes}  where the Bayesian methods together with their corresponding observational inputs are presented. We present our results and conclude with a summary and outlook in the last two sections.

\section{The $\sigma$-$\omega$ model equation of state for neutron star matter}
\label{EoS}

In this study we employed an extended version of the $\sigma $-$\omega$ model to describe the interior of the neutron star \cite{Posfay:2020xgp}. The Lagrangian of the model is given by, 
%
\begin{eqnarray}
\Lagr &=&
%
 \overline{\Psi} \left(
i \slashed{\partial} -m_{N} + g_{\sigma} \sigma -g_{\omega} \slashed{\omega}  + g_{\rho} \slashed{\rho}^{a} \tau_{a}
 \right) \Psi
+ \overline{\Psi}_{e} \left(
i \slashed{\partial} - m_{e}
\right) \Psi_{e}
%
 +\frac{1}{2}\,\sigma \left(\partial^{2}-m_{\sigma}^2 \right) \sigma  \nonumber \\
%
&& - U_{i}(\sigma)- \frac{1}{4}\,\omega_{\mu \nu} \omega^{\mu\nu}+\frac{1}{2}m_{\omega}^2 \, \omega^{\mu}\omega_{\mu} 
%
-\frac{1}{4} \rho_{\mu \nu}^{a} \, \rho^{\mu \nu \, a} + \frac{1}{2} m_{\rho}^2 \, \rho_{\mu}^{a} \, \rho^{\mu \, a} 
 \, ,
\label{eq:wal_lag}
\end{eqnarray}
%
where $\Psi=(\Psi_{n},\Psi_{p})$ is the vector of proton and neutron fields,$m_{N}$, $m_{\sigma}$, $m_{\omega}$ are the nucleon, $\sigma$, and $\omega$  meson masses and $g_{\sigma}$, $g_{\omega}$, and $g_{\rho}$ are the Yukawa couplings corresponding to the $\sigma$-nucleon,  $\omega$-nucleon and $\rho$-nucleon interactions, respectively. The kinetic terms corresponding to the $\omega$ and $\rho$ meson can be written as:
%
\begin{equation}
\omega_{\mu \nu}=\partial_{\mu} \omega_{\nu}-\partial_{\nu} \omega_{\mu} \, , \ \ \textrm{and } \ \
\rho_{\mu \nu}^{a}=\partial_{\mu} \rho_{\nu}^{a} - \partial_{\nu} \rho_{\mu}^{a} + g_{\rho} \epsilon^{abc} \rho_{\mu}^{b} \rho_{\nu}^{c}. \, 
\label{eq:wal_lag2}
\end{equation}
%
In eq.~\eqref{eq:wal_lag} $U_{i}(\sigma)$ is a self interaction term for the $\sigma$-meson and it has the following form:
%
\begin{equation}
\begin{aligned}
U_{34}(\sigma) &=\lambda_{3} \sigma^{3} + \lambda_{4} \sigma^{4}  \, .
\end{aligned}
\label{eq:U_types}
\end{equation}
%
The nuclear interaction  is approximated by the meson fields introduced as in \eqref{eq:wal_lag}. The $\omega$ meson describes the repulsive nature of the interaction while the $\sigma$ meson is responsible for the attraction between the nucleons. The $\rho$ meson takes into account the difference between the proton and the neutron densities. For the description of the neutron rich matter present inside compact stars, protons, electrons and neutrons are considered to be in $\beta$-equilibrium:
%
\begin{equation}
n\, \rightleftharpoons p \, + \, e \,.
\end{equation}
%
This assumption provides a connection between chemical potential of the nucleons and the electrons:
%
\begin{equation}
\begin{split}
\mu_{n} = \mu_{p} + \mu_{e} \, .
\end{split}
\label{eq:asym2}
\end{equation}
%
In this study, we consider the $\sigma $-$\omega$ model in the mean field approximation. This is a reasonable assumption because  quantum fluctuations of nuclear matter do not influence significantly the observable properties of compact stars considering the sensitivity of the currently available measurement techniques \cite{FRG1,FRG2}. Since the temperature of neutron stars is negligible compared to the energy scale corresponding to nuclear matter it is reasonable to assume while calculating the thermodynamics of the system that the nuclear matter is at zero temperature \cite{norman1997compact,Schmitt:2010}. Using these assumptions eq.~\eqref{eq:wal_lag} simplifies considerably. All of the kinetic terms become zero and only the following components of the mesons have non-zero values: $\omega_{0}=\omega$ and $\rho_{0}^{3}=\rho$. With these assumptions the free energy of the model can be calculated as it is described for example in Ref.~\cite{jakovac2015resummation}: 
%
\begin{eqnarray}
f_{T} &=&
%
 f_{F}
\left(
m_{N}-g_{\sigma} \sigma,
\mu_{p} - g_{\omega} \omega + g_{\rho} \rho
\right)
+  f_{F} \left(
m_{N}-g_{\sigma} \sigma,
\mu_{n} - g_{\omega} \omega - g_{\rho} \rho
\right)
+ f_{F} \left(m_{e}, \mu_{e} \right)  \nonumber \\ 
%
&&+ \frac{1}{2} m_{\sigma}^{2} \sigma^2  + U_{i}(\sigma)
%
 - \frac{1}{2} m_{\omega}^2 \omega^2 
%
 - \frac{1}{2} m_{\rho}^2 \rho^2 \, , 
\label{eq:wal_f}
\end{eqnarray}
%
where $\mu_{p}$, $\mu_{n}$ and $\mu_{e}$ are the proton, neutron, and electron chemical potential, respectively. The $f_{F}$ function gives the free energy contribution corresponding to one fermionic degree of freedom. It is given by:
%
\begin{equation}
f_{F}(T,m,\mu)  = -2 T \int \frac{\dd^3 k}{(2 \pi)^3} 
\ln{\left( 1 + \mathrm{e}^{-\beta \left( E_{k}-\mu \right) } \right)}  \ \ \, 
\label{eq:free_en}
\end{equation}
%
where $E_{k}^2  = k^2 + m^2$. As described above in the case of cold nuclear matter of neutron stars one has to take $T \to 0$ in eq.~\eqref{eq:free_en}. This means that the fermionic free energy has only two variables $f_{F}(m, \mu)$.
The couplings in eq.~\eqref{eq:free_en} are determined by nuclear saturation data ~\cite{norman1997compact,meng2016relativistic}.
The values of the nuclear parameters can be found in Table~\ref{table_parameters}.
%
\begin{table}[h]
\caption{\label{table_parameters}Nuclear saturation parameter data}
\begin{center}
\begin{tabular}{ll}
\hline \hline 
\textbf{Parameter}              & \textbf{ Value}        \\
\hline
Binding energy $B$        & $-16.3$ MeV    \\
Saturation density, $n_{0}$    & $0.156$ $ \text{fm}^{-3}$  \\
Nucleon effective mass, $m^{*}$ & $0.6$  $m_{N}$      \\
Nucleon Landau mass $m_{L}$    & $0.83$ $m_{N}$ \\
Compressibility, $K_0$      & $240$ MeV   \\ 
(A)symmetry energy, $S_{0}$      & $32.5$ MeV   \\   
\hline \hline 
\end{tabular}
\end{center}
\end{table}
%
The definition of the Landau mass is given by (based on Ref.~\cite{norman1997compact}):
\begin{table}[h]
\begin{center}
\begin{tabular}{c|ccccc}
\hline \hline 
Model &	$n_0$ &	 $B$ & 	$K$ &$S_0$ & $m^{*}$ \\
&					[fm$^{-3}$] &			[MeV]  &			[MeV]  &		[MeV] & 		[$m_N$] \\
\hline
NL$\rho$ 	&0.1459 &-16.062 &203.3 & 30.8 & 0.603 \\
NL$\rho$$\delta$ 	&0.1459 &-16.062 &203.3 & 31.0 & 0.603 \\
DBHF 	&0.1810 &-16.150 &230.0 & 34.4 & 0.678 \\
DD 		&0.1487 &-16.021 &240.0 & 32.0 & 0.565 \\
D3C 	&0.1510 &-15.981 &232.5 & 31.9 & 0.541 \\
KVR 	&0.1600 &-15.800 &250.0 & 28.8 & 0.805 \\
KVOR 	&0.1600 &-16.000 &275.0 & 32.9 & 0.800 \\
DD-F 	&0.1469 &-16.024 &223.1 & 31.6 & 0.556 \\
\hline
\end{tabular}
\end{center}
\caption{\label{table_realistic_EoS}Nuclear saturation parameters for several EoS models presented in~\cite{Klahn:2006ir}.}
\end{table}
\begin{equation}
\begin{aligned}
m_{L} &=\frac{k_{F}}{v_{F}}  \quad \text{with} \quad
v_{F} &=\left.\frac{\partial E_{k}}{\partial k} \right|_{k=k_{F}} \, .
\end{aligned}
\label{eq:landau_mass}
\end{equation}
%
where $k=k_{F}$ is the Fermi-momentum and $E_{k}$ is the dispersion relation corresponding to the nucleons. The Landau mass is not independent of the effective mass in the mean field approximation. Substituting the energy of nucleons from eq.~\eqref{eq:free_en} into the definition of the Landau mass yields the above relation: 
%
\begin{equation}
\begin{split}
m_{L}= \sqrt{k_{F}^2 + m_{N, eff}^2} \, .
\end{split}
\label{eq:effmass_vs_landau_mass}
\end{equation}
%
This connection makes it impossible to fit the value of the Landau mass and the effective mass simultaneously \cite{meng2016relativistic}. One of them has to be chosen as a free parameter and only after fixing its value the other one becomes completely determined. The compression modulus is defined as usual as in for example in Refs.~\cite{Schmitt:2010,norman1997compact}:
%
\begin{equation}
\begin{split}
K =k_{F}^2 \frac{\partial^2 }{\partial k_{F}^2} \left( \frac{\epsilon}{n} \right)
= 9 n^2 \frac{\partial^2}{\partial n^2} \left( \frac{\epsilon}{n} \right) \, .
\end{split}
\label{eq:K}
\end{equation}
%
The asymmetry energy is defined as it is described for example in \cite{norman1997compact}: 
%
\begin{equation}
a_{sym} = \frac{1}{2} \left. 
\frac{\partial^2 }{\partial t^2} \left( \frac{\epsilon}{n} \right) \right|_{t=0} \,.
\label{eq:asym_def}
\end{equation}
%
Here $t=\frac{n_{n}-n_{p}}{n_{B}}$ is the relative difference between the number of neutrons and protons.

To describe the crust of the neutron star we complemented the extended $\sigma$-$\omega$ model with a low density EoS. The two EoS are joined at the point where both models hold the same pressure. For the crust EoS we used the widely known BPS equation of state \cite{BPS}. 

In this study our aim is to use astrophysical data to estimate the most relevant parameters of the model, some of them already measured in the laboratory. Earlier works suggest the nuclear parameters values that we present in Table~\ref{table_parameters} whereas Table~\ref{table_realistic_EoS} shows parameter values for other EoS models for the sake of comparison and also shows the parameter similarity of our model with other relativistic field models like DD. The Landau mass and the compression modulus have the largest influence on neutron star observables \cite{Posfay:2020xgp,symwal}. To provide data for the Bayesian analysis we fitted the extended $\sigma$-$\omega$ model using different values for the above mentioned two parameters at saturation density, $K_0 \equiv K(n_0)$ and $S_0 \equiv S(n_0) \equiv a_{sym}(n_0)$. Using their values in Table~\ref{table_parameters} as reference points we varied them within a 80\% to 110\% range. Thus, we have chosen 10 points within the above mentioned range for each parameter and ended up with 1000 EoS models taking into account every possible combination. 

\section{Neutron Star Sequences}
\label{Sequences}

Predictions for the properties of static neutron stars from the extended $\sigma$-$\omega$ EoS sets are obtained by solving the Tolman\,--\,Oppenheimer\,--\,Volkoff equations~\cite{Tolman:1939jz,Oppenheimer:1939ne}:
 \begin{eqnarray}
 \label{TOV}
\frac{\dd P( r)}{\dd r}&=& 
-\frac{\left(\varepsilon( r)+P( r)\right)
\left(m( r)+ 4\pi r^3 P( r)\right)}{r\left(r- 2m( r)\right)},\\
\frac{\dd m( r)}{\dd r}&=& 4\pi r^2 \varepsilon( r).
\label{eq:TOVb}
 \end{eqnarray}
from which the mass is derived.  The above equations are to be solved  simultaneously with the neutron star EoS: $p(\varepsilon)$. The radial functions $\varepsilon(r)$ and $p(r)$ represent the energy density and pressure in the interior of the star. The boundary conditions for the system are zero pressure at the star surface
$p(r=R)=0$ as well as null mass at the origin $m(r=0)=0$. The total mass $M$ is defined as $M=m(r=R)$. To determine the mass and radius of a single star it is necessary to specify its central density $\varepsilon_c$ at the origin. To obtain the entire sequence of neutron stars displayed in the corresponding $M$-$R$ diagram, $\varepsilon_c$ is increased starting from a low density value around nuclear saturation. In addition, the dimensionless tidal deformability of neutron stars $\Lambda$ is computed following the approach introduced in~\cite{Hinderer:2007mb}. It is related to the Love number $k_2$:
\begin{equation} 
\Lambda = \frac{2}{3} \frac{ R^{5}}{M^{5}} k_2
\end{equation}
whose computation involves a perturbation treatment of the static, spherical metric of the neutron star, see~\cite{Damour:2009vw,Binnington:2009bb,Yagi:2013awa,Hinderer:2009ca} for details.

\section{Bayesian inference Formalism}
\label{Bayes}

Let's introduce a notation for a vector of parameters, so that each vector represents one model from the considered EoS:
\begin{equation}
\label{pi_vec}
\overrightarrow{\pi}_q = \left\{{m_L}_{(i)},{K_0}_{(j)},{S_0}_{(k)}\right\},
\end{equation}
where $q = 0\dots N-1$
(with $N = N_1\times N_2\times N_3$) as $q = N_2\times N_3\times i + N_3\times j + k$ and
$i = 0\dots N_1-1$, $j = 0\dots N_2-1$, $k = 0\dots N_3-1$, here $N_1$, $N_2$ and $N_3$ denote the cardinalities of the ordered sets of values of the parameters $m_L$, $K_0$ and $S_0$ respectively.
The full likelihood for a given $\overrightarrow{\pi}_q$ can be calculated as a product of all likelihoods, since the considered constraints are independent of each other
\begin{equation}
\label{eq:p_event}
P\left(E\left|\overrightarrow{\pi}_{q}\right.\right)= \prod_{w} P\left(E_{w}\left|\overrightarrow{\pi}_{q}\right.\right).
\end{equation}
In the equation above $w$ is an index for the constraints. The posterior distribution is given by Bayes theorem
\begin{equation}
\label{eq:bayes}
P\left(\overrightarrow{\pi}_{q}\left|E\right.\right)=\frac{P\left(E\left|\overrightarrow{\pi}_{q}\right.\right)P\left(\overrightarrow{\pi}_{q}\right)}{\sum\limits _{p=0}^{N-1}P\left(E\left|\overrightarrow{\pi}_{p}\right.\right)P\left(\overrightarrow{\pi}_{p}\right)},
\end{equation}
where $P\left(\overrightarrow{\pi}_{q}\right)$ is a prior distribution of a model taken to be uniform: $P\left(\overrightarrow{\pi}_{q}\right)=1/N$. 

\subsection{Likelihood of a model for the GW170817 ($\Lambda_1$--$\Lambda_2$ constraint)}
\label{bayes_1}
In order to implement the tidal deformability constraint on the compact star EoS, reflected on the $\Lambda_1$--$\Lambda_2$ diagram that includes probability regions from GW170817 event~\cite{TheLIGOScientific:2017qsa,Abbott:2018exr}, we employ the following formula for the likelihood:
\begin{equation}
\label{eq:lhoodLL}
P\left(E_{GW}\left|\pi_q\right.\right) = \int_{l} \beta\left(\Lambda_1(n_c), \Lambda_2(n_c)\right)\dd{n_c}
\end{equation}
where $l$ is the length of the line on the $\Lambda_1$--$\Lambda_2$ diagram produced by $\overrightarrow{\pi_q}$, and $n_c$ is the central density of a star.  $\beta(\Lambda_1, \Lambda_2)$ is the Probability Distribution Function (PDF) that has been reconstructed (as previously done in~\cite{Ayriyan:2018blj}) by the method of Gaussian kernel density estimation using $\Lambda_1$--$\Lambda_2$ data available in the LIGO web-page \cite{LIGO}.

\subsection{Likelihood of a model for the maximum mass (maximum mass constraint)}
\label{bayes_2}
The likelihood of a model is the conditional probability of the maximum mass under the considered maximum mass constraints. Here we have considered the recent estimation of the mass of the heaviest known pulsar {PSR\,J0740+6620} given by the measurements presented in~\cite{Cromartie:2019kug} $2.14_{-0.09}^{+0.10}~\textrm{M}_{\odot}$. Since it has relatively large error deviation, the mass estimation of {PSR\,J0348+0432}  $2.01_{-0.04}^{+0.04}~\mathrm{M_{\odot}}$~\cite{Antoniadis:2013pzd} has been considered as well. The last one is lower, but it has a tiny standard deviation of the error, so that it helps to constrain the maximum mass from below starting from the $3\sigma$ region. Additionally, the constraint for the upper limit on the maximum mass $2.16_{-0.15}^{+0.17}~\textrm{M}_{\odot}$~\cite{Rezzolla:2017aly} has been added to the analysis. Note, that all those values are given at the $68.3\%$ confidence level. The likelihood for the maximum mass constraint has been introduced in the following form:
\begin{equation}
\label{eq:lhoodMass}
P\left(E_{M}\left|\pi_q\right.\right) = \Phi(M_q, \mu_\mathrm{C}, \sigma_\mathrm{C})\times\Phi(M_q, \mu_\mathrm{A}, \sigma_\mathrm{A}) \times\left(1-\Phi(M_q, \mu_\mathrm{U}, \sigma_\mathrm{U})\right),
\end{equation}
where $M_q$ is the maximum mass for a vector of parameters $\pi_q$, and $\Phi(M, \mu, \sigma)$ is the Cumulative Distribution Function (CDF) of the standard normal distribution.
The mass measurement for {PSR\,J0740+6620} has been approximated with $(\mu_C = 2.14, \sigma_C = 0.09)$. For {PSR\,J0348+0432} the original mean $\mu_\mathrm{A}=2.01$ and deviation $\sigma_\mathrm{A}=0.04$ have been used. The upper limit constraint has been approximated by $(\mu_\mathrm{U} = 2.16, \sigma_\mathrm{U} = 0.17)$.


\subsection{Likelihood of a model for mass and radius  ($M$-$R$ constraint)}
\label{bayes_3}
The results of the Neutron Star Interior Composition Explorer (NICER) observation of {PSR\,J0030+0451} have been recently reported in a collection of publications, see for instance Ref.~\cite{Miller:2019cac,Riley:2019yda}. There were two estimates of the mass and equatorial radius based on mutually exclusive assumptions about the uniform-temperature emitting spots. The first  radius and mass estimates are  $M_1={1.44}_{-0.14}^{+0.15}\,{M}_{\odot }$ and  $R_1={13.02}_{-1.06}^{+1.24}\,\mathrm{km}$~\cite{Miller:2019cac} whereas the second estimates are $M_2=1.34_{-0.16}^{+0.15}\,{M}_{\odot }$ and $R_2={12.71}_{-1.19}^{+1.14}\,\mathrm{km}$~\cite{Riley:2019yda}.
Here we introduced approximations of those estimates by considering a bivariate normal distribution PDF exactly as in~\cite{Blaschke:2020qqj}. Assuming that those two estimates are equiprobable, since they are mutually exclusive, a likelihood for the $M$-$R$ constraint has been introduced as follows
\begin{eqnarray}
\label{eq:lhoodMR}
P\left(E_{MR}\left|\pi_q\right.\right) = 0.5\int_{l} \mathcal{N}\left(\mu_M^{(1)},\,\sigma_M^{(1)},\,\mu_R^{(1)},\,\sigma_R^{(1)},\alpha^{(1)}\right)\dd{n_c} \nonumber\\
+\,0.5\int_{l} \mathcal{N}\left(\mu_M^{(2)},\,\sigma_M^{(2)},\,\mu_R^{(2)},\,\sigma_R^{(2)},\alpha^{(2)}\right)\dd{n_c}, 
\end{eqnarray}
where $\left(\mu_M, \mu_R\right)$ is the mathematical expectation in the $M$-$R$ plot,  $\sigma_M$ and $\sigma_R$ are standard  deviations for mass and radius respectively, and $\rho$ is the rotation angle of an ellipse in Fig.~\ref{MR_constraints} which corresponds to correlation  between mass and radius. Here $l$ is the length of the curve on $M$-$R$ plot for a particular EoS and $n_c$, as mentioned above, is the central density of a star.
The approximations of \cite{Miller:2019cac} and \cite{Riley:2019yda} by bivariate normal distributions are 
\begin{equation}
\{ \mu_M^{(1)} = 1.44\,\textrm{M}_\odot, \,\sigma_M^{(1)} = 0.145\,\textrm{M}_\odot, \,\mu_R^{(1)} = 13.02\,\textrm{km}, \,\sigma_R^{(1)} = 1.150\,\textrm{km}, \,\alpha^{(1)} = 7.038^\mathrm{o} \}, \nonumber
\end{equation}
\begin{equation}
\{ \mu_M^{(2)} = 1.34\,\textrm{M}_\odot, \,\sigma_M^{(2)} = 0.155\,\textrm{M}_\odot, \,\mu_R^{(2)} = 12.71\,\textrm{km}, \,\sigma_R^{(2)} = 1.165\,\textrm{km}, \,\alpha^{(2)} = 7.359^\mathrm{o} \}. \nonumber
\label{normal}
\end{equation}
\subsection{Marginalization}
\label{bayes_4}
For the convenience of analysis of the Bayesian inference results, the marginalisation procedure has been used. In order to the show posterior distribution of one parameter, namely marginalization over other parameters, it is required that:
\begin{eqnarray}
\label{eq:marg1D}
P\left({m_L}_{(i)}\left|E\right.\right)=\sum_{j,k}P\left({m_L}_{(i)},{K_0}_{(j)},{S_0}_{(k)}\left|E\right.\right).
\end{eqnarray}
The 2D posterior distribution is obtained by a subsequent marginalization over the remaining one parameter:
\begin{eqnarray}
\label{eq:marg2D}
P\left({m_L}_{(i)},{K_0}_{(j)}\left|E\right.\right)=\sum_{k}P\left({m_L}_{(i)},{K_0}_{(j)},{S_0}_{(k)}\left|E\right.\right).
\end{eqnarray}
\begin{figure}[htpb!]
\centering
\resizebox{0.75\textwidth}{!}{
  \includegraphics{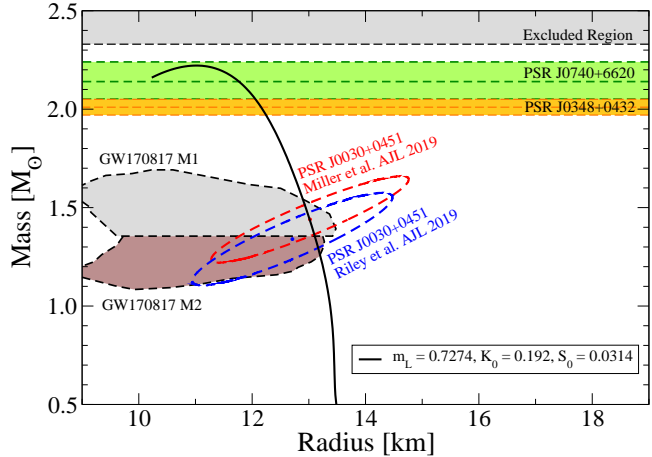}
}
\caption{Mass-radius diagram for neutron stars. Shaded regions correspond to the different measurements used as inputs for the Bayesian analysis. The two horizontal bands correspond to the pulsar masses M=$2.14_{-0.09}^{+0.10}~\textrm{M}_{\odot}$ of {PSR\,J0740+6620}~\cite{Cromartie:2019kug}, and M=$2.01_{-0.04}^{+0.04}~\mathrm{M_{\odot}}$ of {PSR\,J0348+0432}~\cite{Antoniadis:2013pzd}. The grey and brown regions above and below  the mass value of M=$1.365~\textrm{M}_{\odot}$ correspond to the mass-radius estimates for the two compact stars of the merger in GW170817~\cite{TheLIGOScientific:2017qsa,Abbott:2018exr}.  The elliptical regions denoted by dashes lines correspond to the mass and radius measurement of the pulsar {PSR\,J0030+0451}: M=${1.44}_{-0.14}^{+0.15}\,${M}$_{\odot }$ with  R=${13.02}_{-1.06}^{+1.24}\,\mathrm{km}$~\cite{Miller:2019cac} or M=$1.34_{-0.16}^{+0.15}\,${M}$_{\odot }$ and R=${12.71}_{-1.19}^{+1.14}\,\mathrm{km}$~\cite{Riley:2019yda}. The forbidden upper region above $M>2.33 $M$_{\odot}$ has been derived from GW170817 together with the associated kilonova AT2017gfo in Ref.~\cite{Most:2018hfd}.
}
\label{MR_constraints}       
\end{figure}
Fig.~\ref{MR_constraints} shows the input constraints for our Bayesian analysis in the $M - R$ diagram of neutron stars where also one of the most probable EoS sequence is displayed for completeness and clarity of our results.  
%
\section{Results}
\label{Results}

\begin{figure*}[!htb]
\begin{centering}
$\begin{array}{cc}
\includegraphics[width=0.5\textwidth]{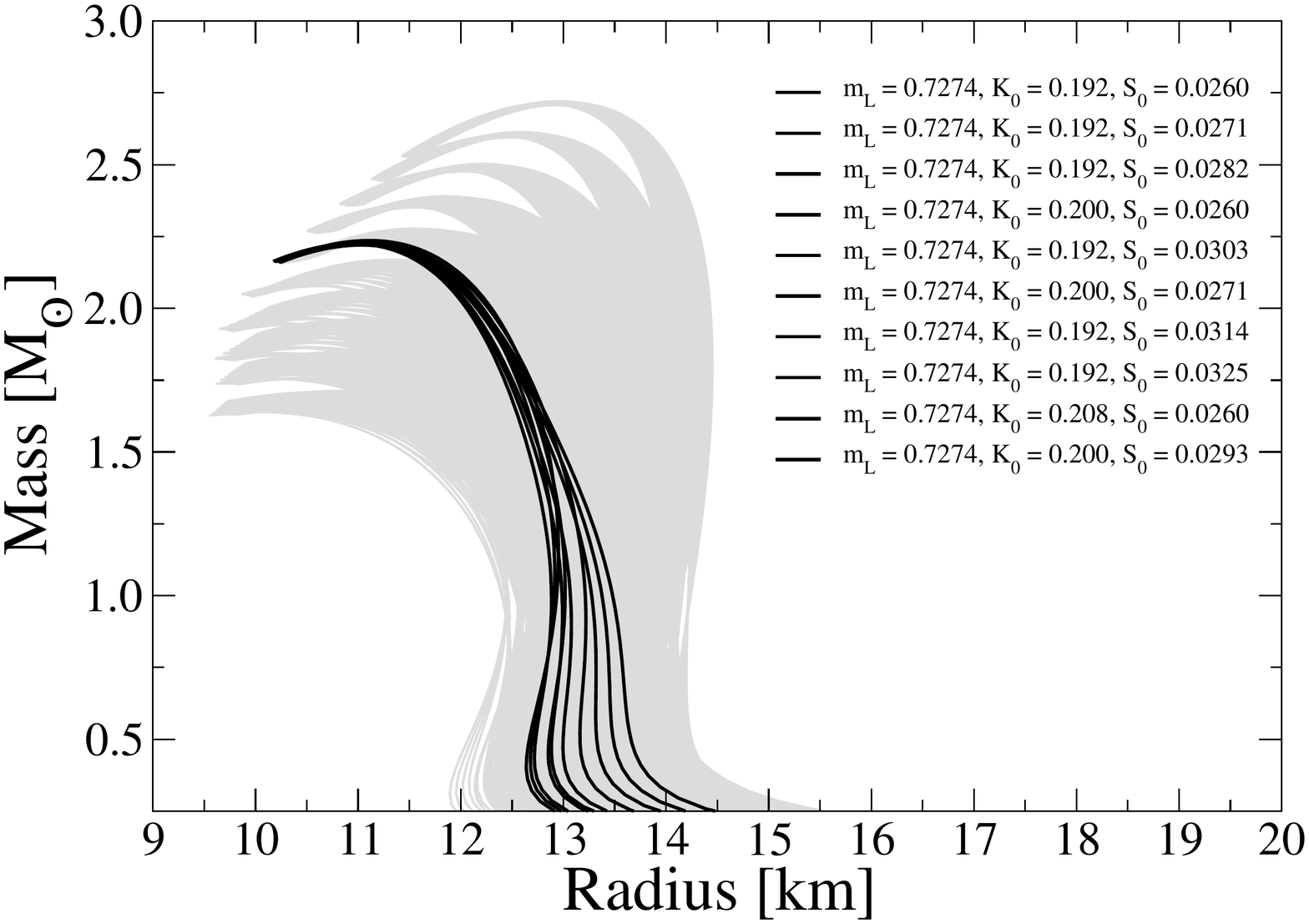} & \hspace{-0.5cm} \includegraphics[width=0.5\textwidth]{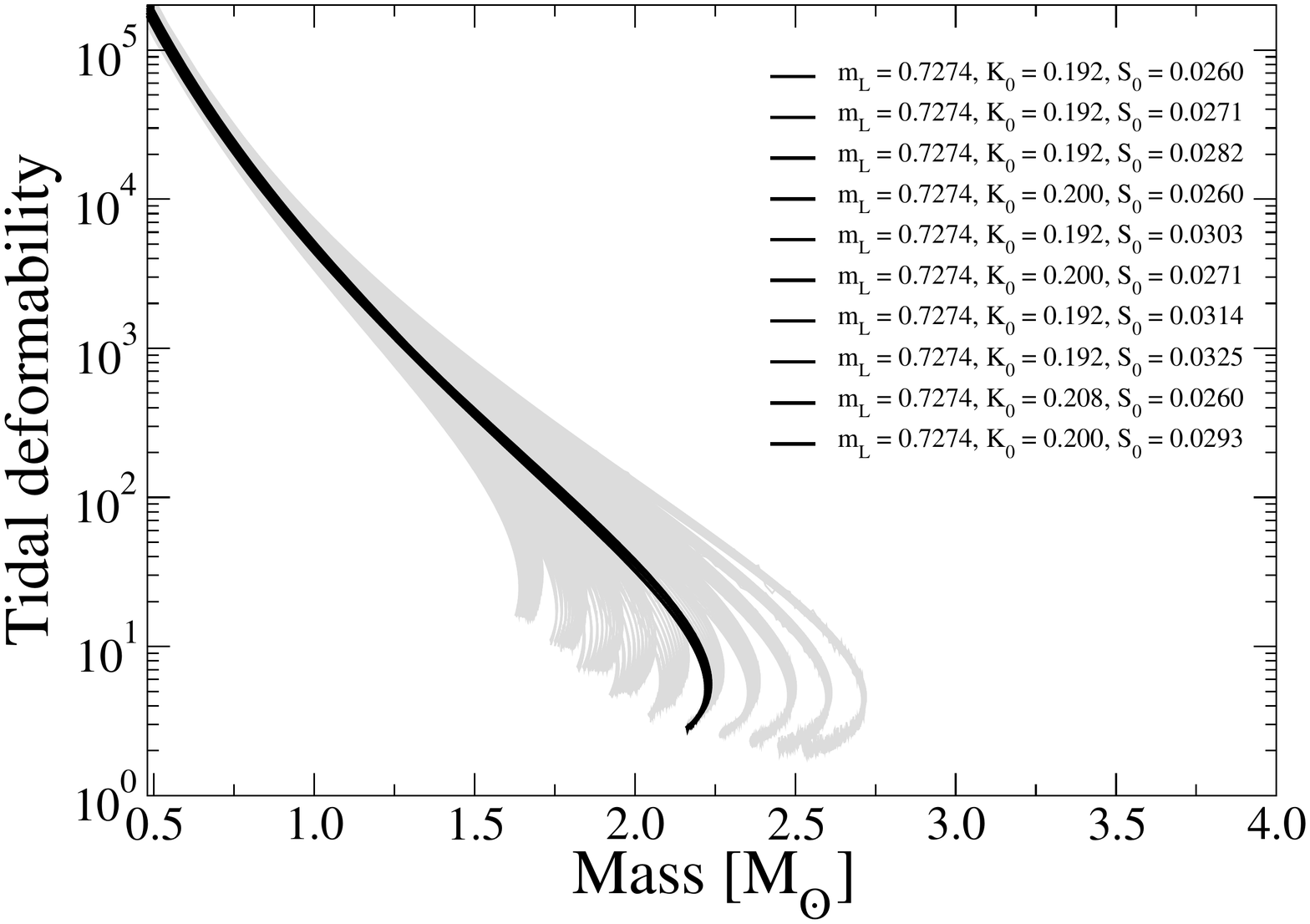} \\
\includegraphics[width=0.5\textwidth]{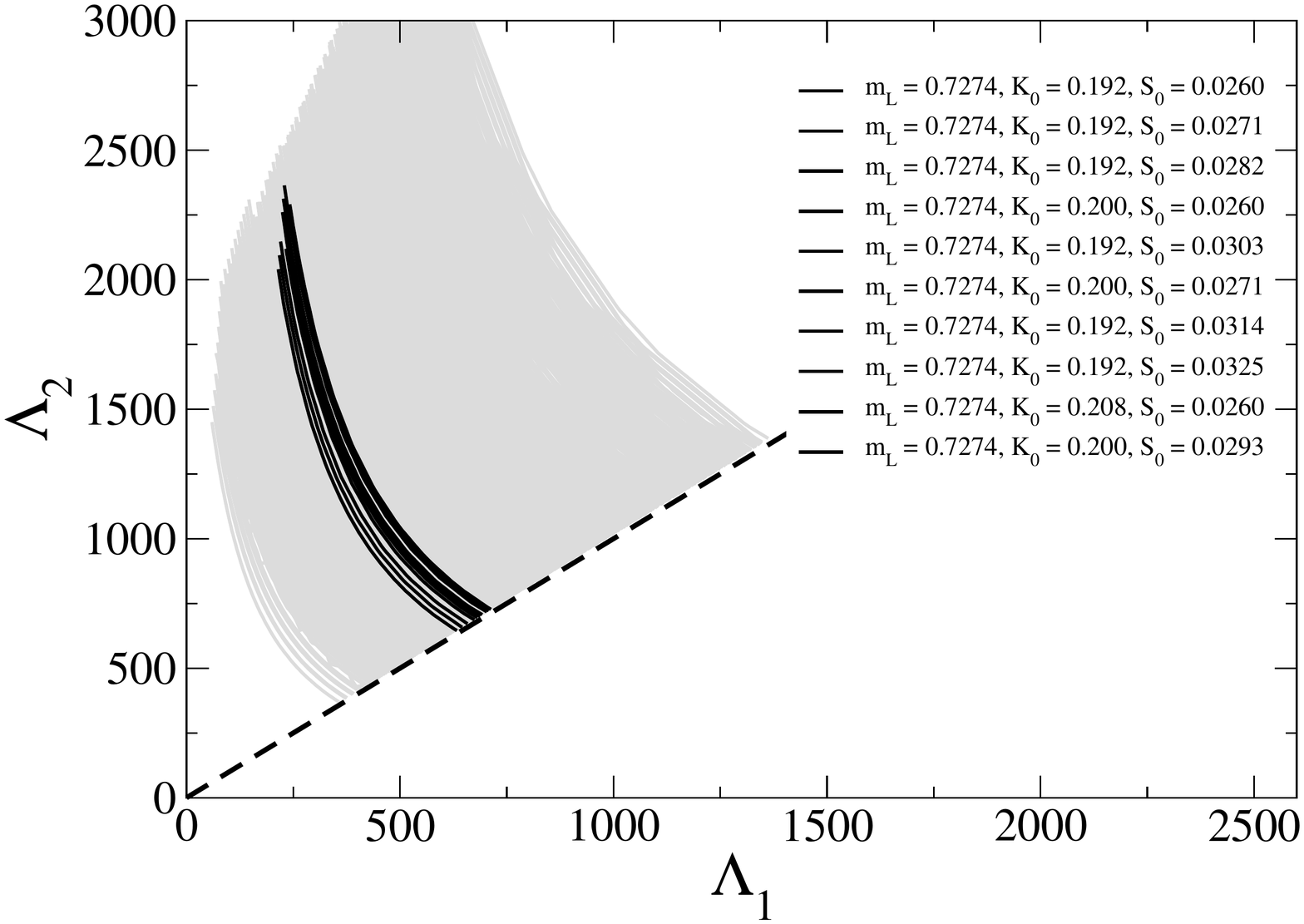}  & \hspace{-0.5cm}  \includegraphics[width=0.5\textwidth]{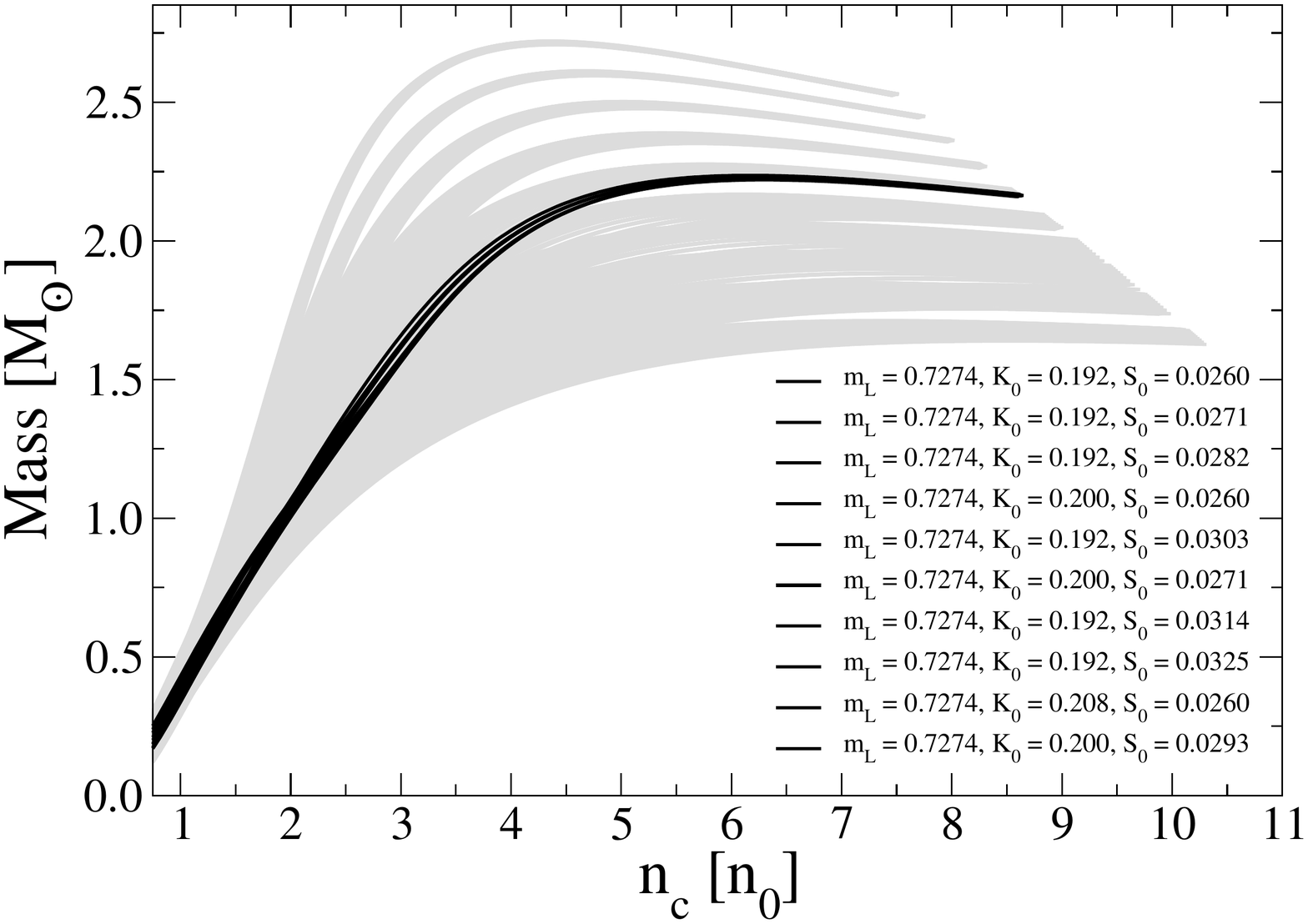} \\
\end{array}$ 
\par\end{centering}
\caption{\label{sequences} Neutron star sequences from the extended $\sigma$-$\omega$ EoS model. The four diagrams relate the neutron star measurements that are used as input for our Bayesian analysis. The dark lines correspond to the highest posterior probability neutron star sequences resulting from the Bayesian method.}
\end{figure*}

\begin{figure*}[!htb]
\begin{centering}
$\begin{array}{ccc}
\includegraphics[width=0.35\textwidth]{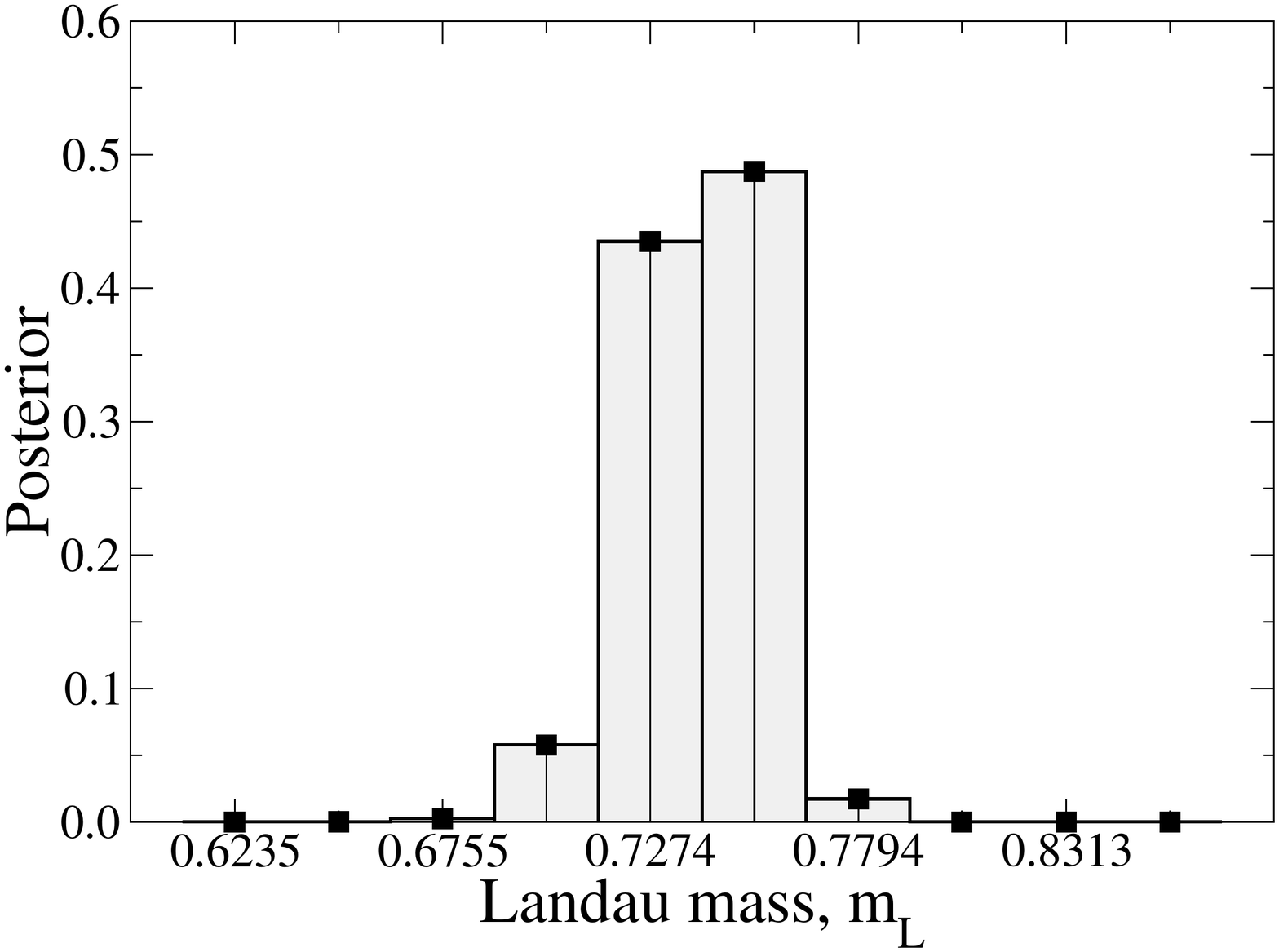}  & \hspace{-0.5cm} \includegraphics[width=0.35\textwidth]{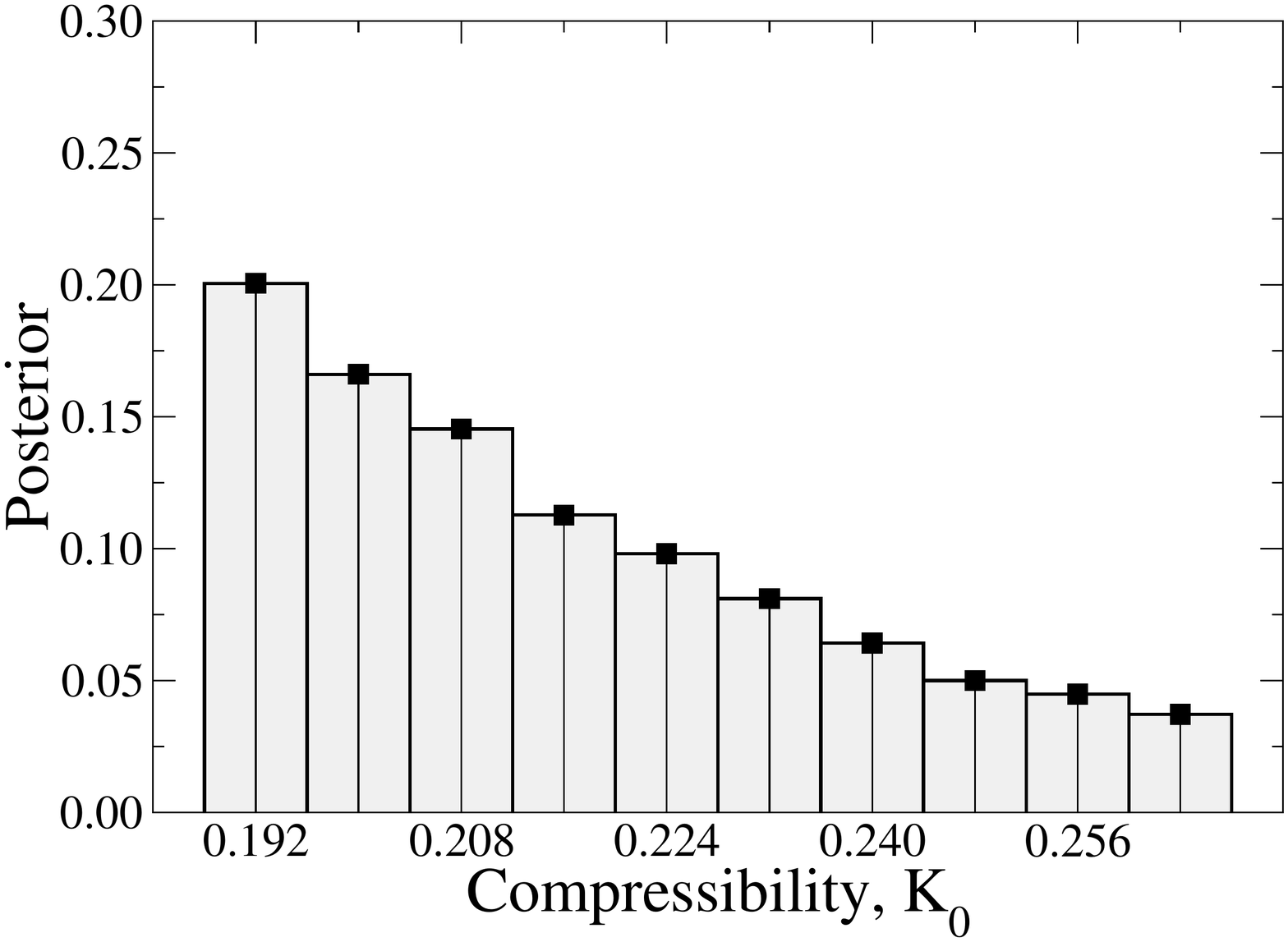} & \hspace{-0.5cm} \includegraphics[width=0.35\textwidth]{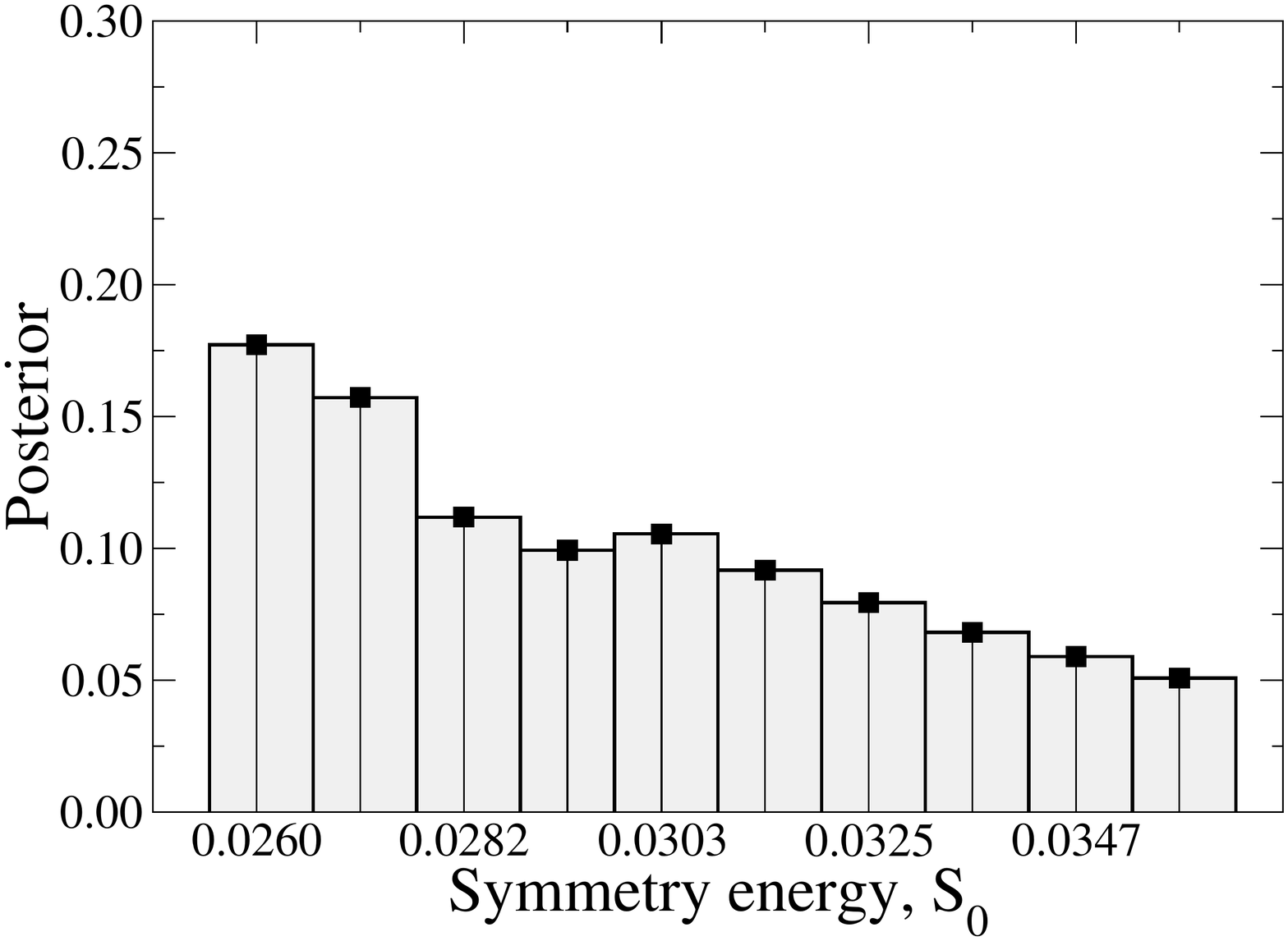} \\
\includegraphics[width=0.35\textwidth]{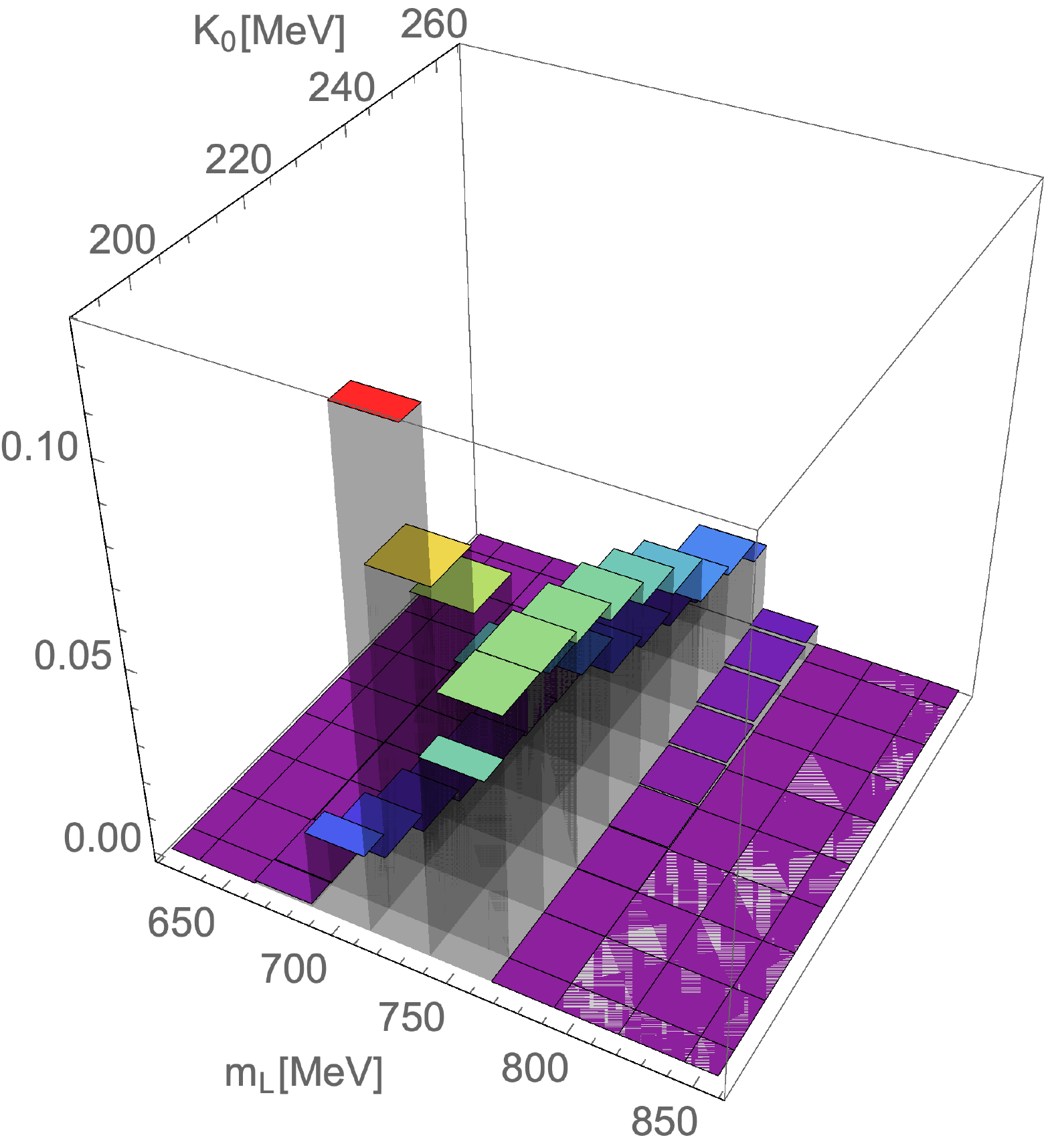}  & \hspace{-0.5cm} \includegraphics[width=0.35\textwidth]{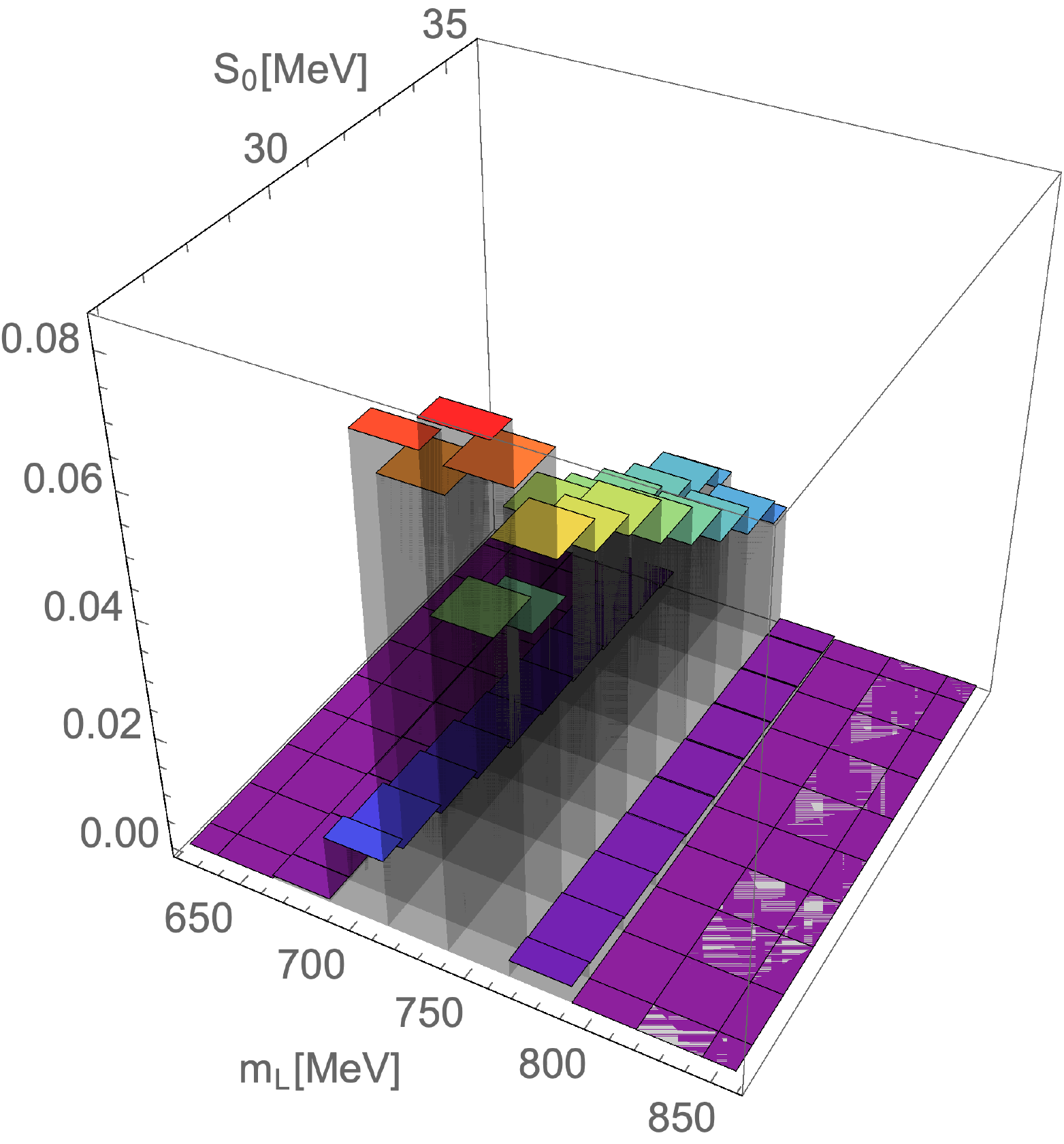} & \hspace{-0.5cm} \includegraphics[width=0.35\textwidth]{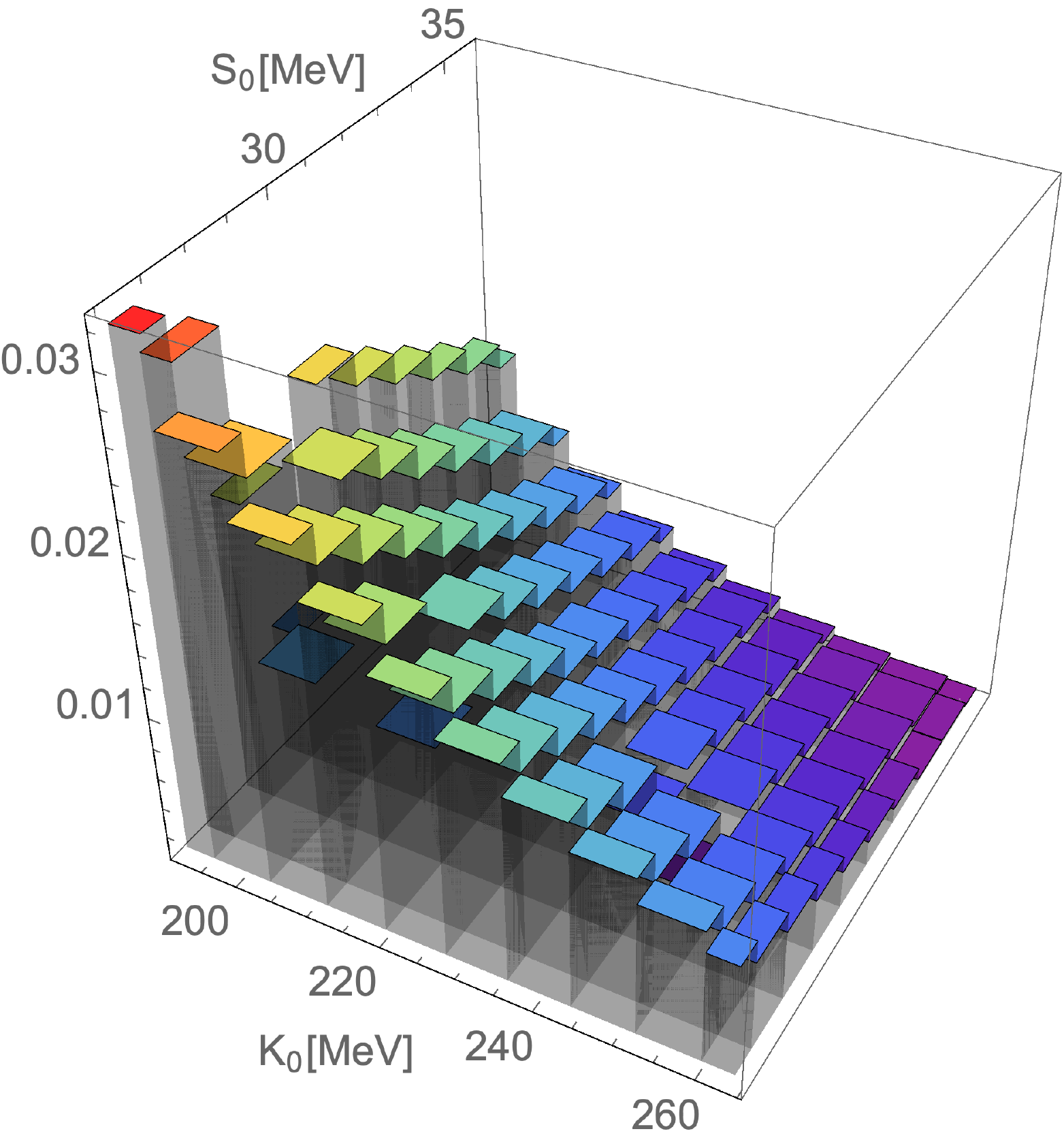} \\
\end{array}$ 
\par\end{centering}
\caption{\label{Bayes_three_parameters} Posterior probabilities of the Landau mass $m_L$, symmetry energy $S_0$ and the nuclear compressibility $K_0$ at saturation resulting from a Bayesian analysis. The upper figures represent the marginalised probabilities, while the lower ones correspond to two-parameter probability distributions marginalised over a third one.}
\end{figure*}

\begin{figure*}[!htp]
\begin{centering}
$\begin{array}{cc}
\includegraphics[width=0.5\textwidth]{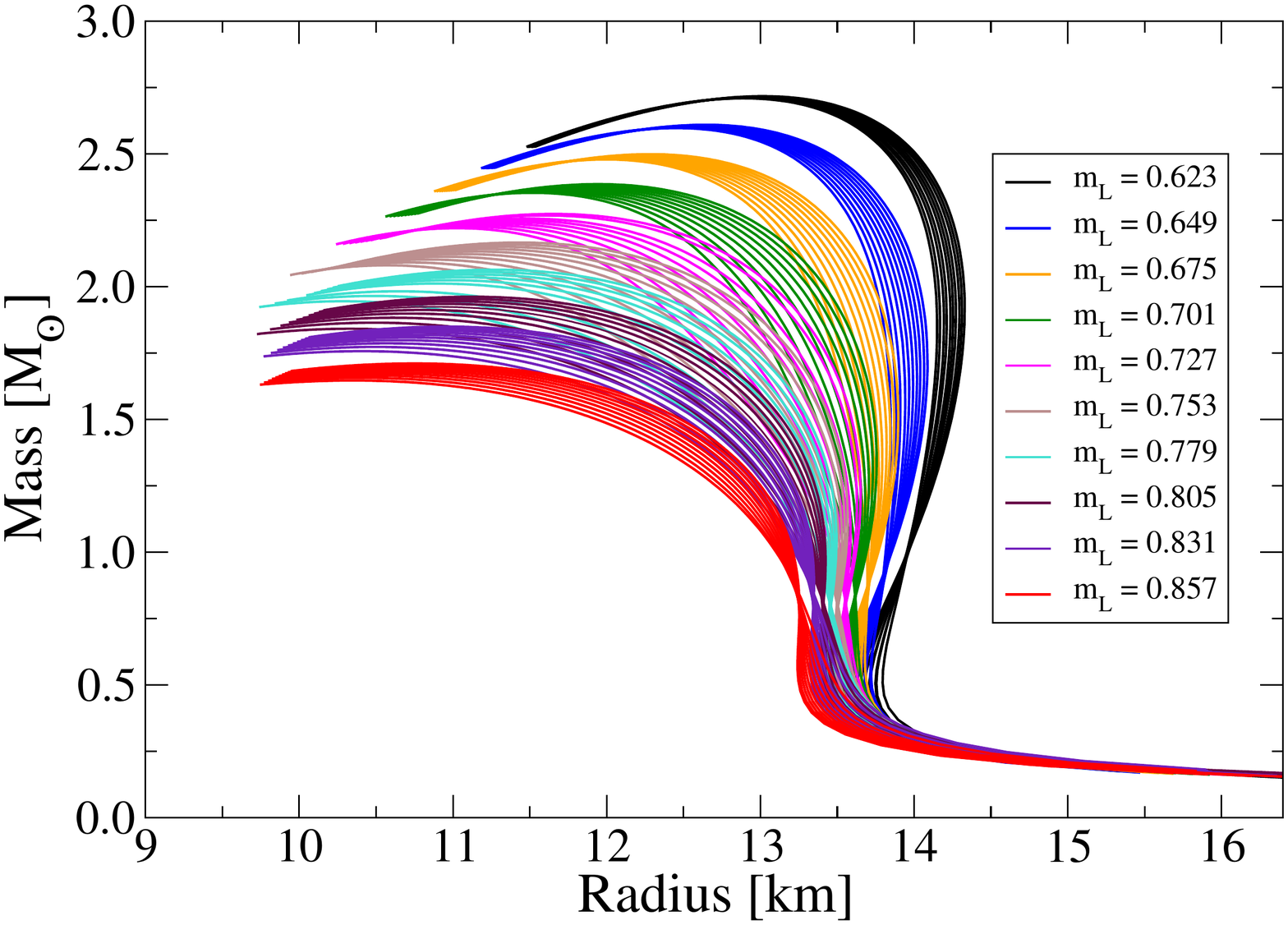}  & \hspace{0cm} \includegraphics[width=0.5\textwidth]{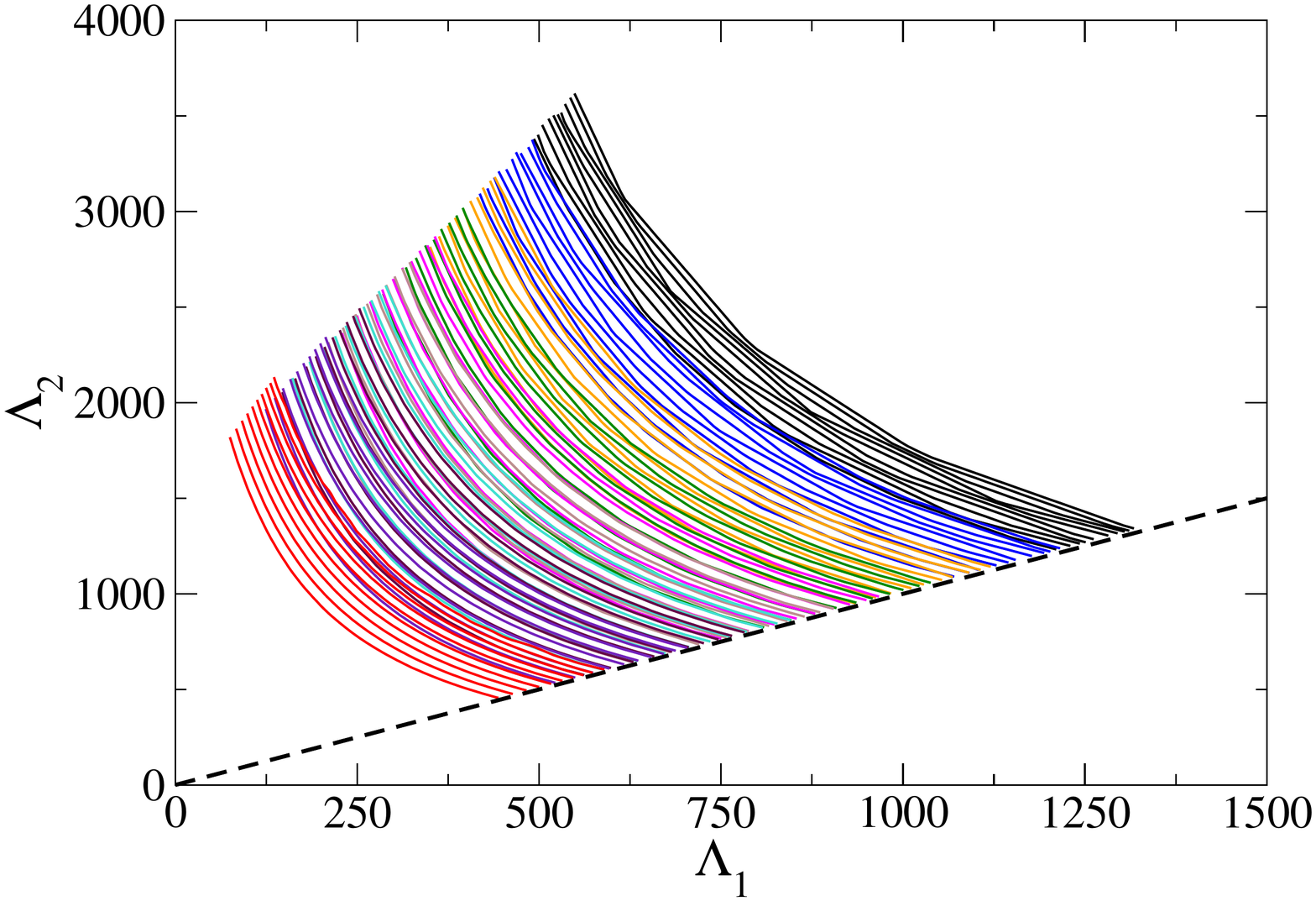} \\
\includegraphics[width=0.55\textwidth]{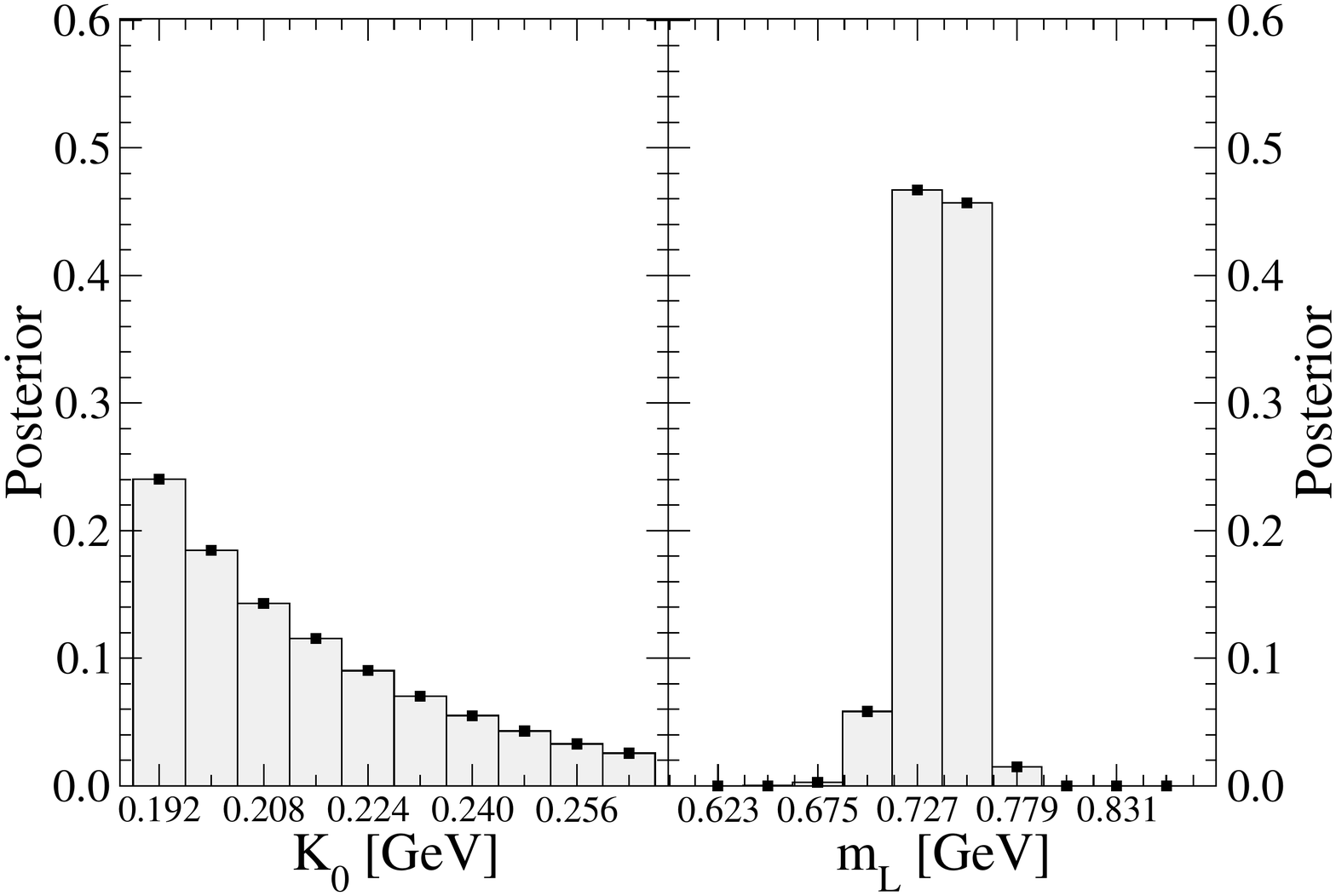}  & \hspace{0cm} \includegraphics[width=0.45\textwidth]{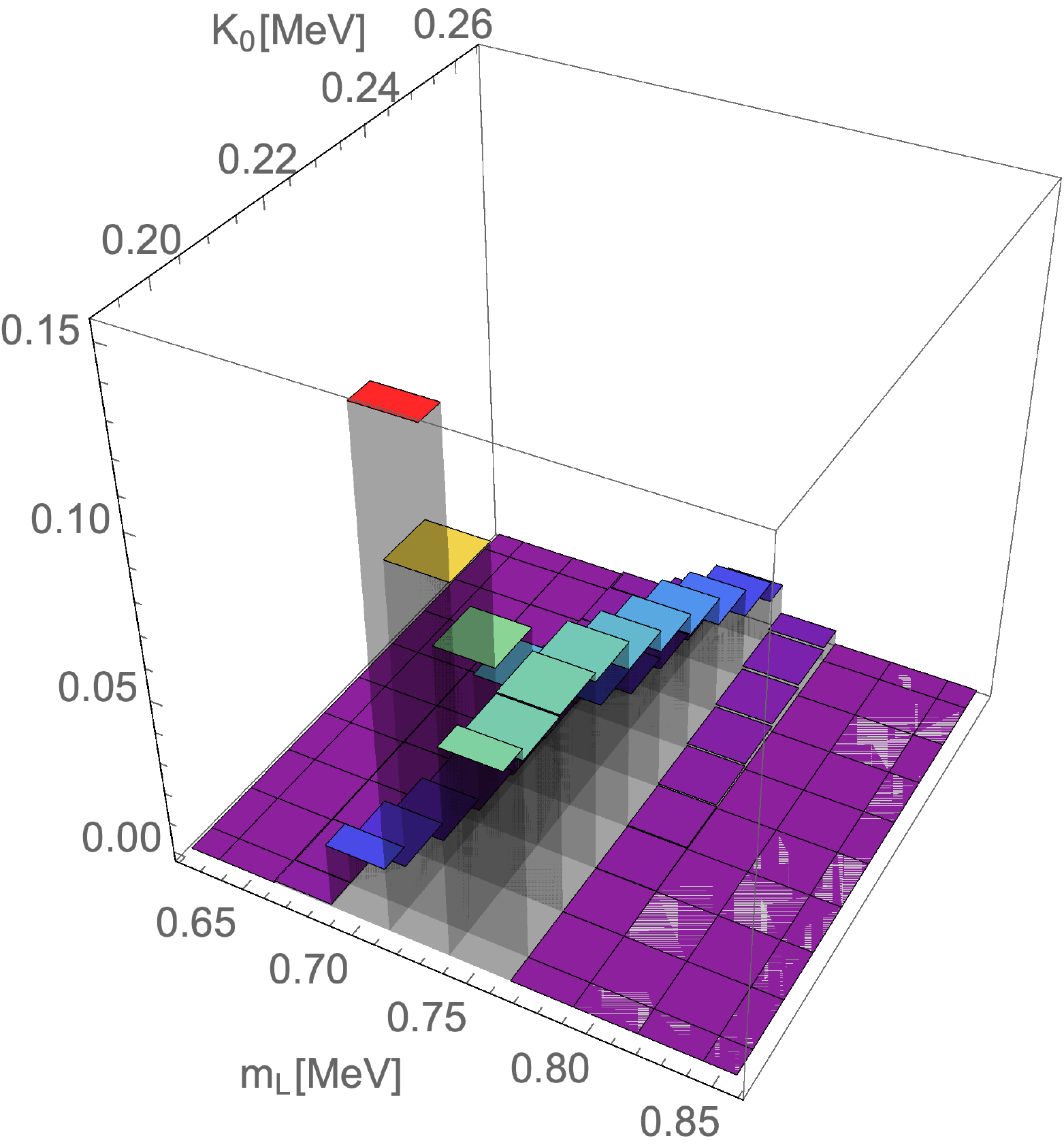} \\
\end{array}$ 
\par\end{centering}
\caption{\label{Bayes_two_parameters}Posterior probabilities of the Landau mass $m_L$ and the nuclear compressibility $K_0$ at saturation resulting from a Bayesian analysis with a fixed symmetry energy value $S_0=0.0325$ GeV. 
The upper figures show the sequences of neutron stars in the mass-radius  and lambda diagrams, respectively. Groups of the same color style share the same Landau mass $m_L$ for 10 values of the nuclear compressibility $K_0$.
The lower figures show the results of a two-parameter posterior probability distribution. The estimated the Landau mass is $738\pm17$ MeV which coincides with the result for the three parameter Bayesian analysis.}
\end{figure*}

In order to cover a relevant set of compact star configurations we have varied the compressibility parameter in the range $K_0$[GeV] $ =0.192, 0.2, 0.208, ...,0.264$ as well as the symmetry energy in the range $S_0$[GeV]$=0.026, 0.028, 0.30, ...,0.036$ and the Landau mass range  $m_L$[GeV]$=0.623, 0.649, 0.675, ...,0.857$. The neutron star sequences are presented in Fig.~\ref{sequences}, where several plots that relate mass, radius, central density and tidal deformability for the entire set of EoS models are included. As it can be seen in the $M$-$R$ diagram of Fig.~\ref{sequences},  we find that the resulting neutron star sequences cover radius values of between 11 km and 14.5 km,  for masses above half a solar mass and a maximum mass value of about 2.7 M$_{\odot }$ for the stiffest EoS. It can also be seen that matter can be compressed up to 6 times saturation density for the softer EoS which are unable to support the observed 2 M$_{\odot}$ neutron star.
In all the plots but the  $\Lambda_{1}-\Lambda_{2}$  diagram, sequences of neutron stars appear to be grouped in sets being separated by gap regions. The sequences in these groups share the same values of $m_L$, an observation that points at the influence of the Landau mass over the other two parameters, especially for the highest masses. The black line sequences represent the top 10 EoS models with the highest probabilities with respect to the rest which has been greyed out. Fig.~\ref{Bayes_three_parameters} shows the posterior probabilities resulting from our Bayesian analysis. The upper figures show the marginalised probabilities for each of our EoS parameters under study (see Eq.~\ref{normal}) whereas the lower LEGO plots represent the 2D posterior distributions for each pair of parameters that have been marginalized over a third one. These probability regions correspond to the two parameters $K_0$ and $S_0$ with marginalization over $m_L$. 
The effects of the EoS parameters on the maximum mass have been found to follow a variation within one
order of magnitude as follows~\cite{Posfay:2020xgp,Barnafoldi:2020}
\begin{equation}
\Delta M_{max}(\delta m_L) \overset{10 \times}{ >} \Delta
M_{max}(\delta K_0) \overset{10 \times}{ >}  \Delta M_{max}(\delta
S_{0}).
\end{equation}
Most importantly, in order to corroborate the major effect of $m_L$ over the remaining parameters on neutron stars we have reduced our Bayesian study to two EoS parameters, keeping the symmetry energy fixed. Therefore, we vary the nuclear compressibility and the Landau mass because these quantities significantly contribute to the stiffness of the EoS whereas the symmetry energy has a major influence on the stellar radius. We present the setup and results in Fig.~\ref{Bayes_two_parameters} where the $M$-$R$ diagram clearly shows the subsets of sequences with the same $m_L$ value. The posterior probabilities are displayed in the lower plots which indicate the same value for the  $m_L = 739\pm17$ MeV which coincides with the previous result from the three EoS parameter study as well as being compatible with the one parameter study in Ref.~\cite{Alvarez-Castillo:2020aku}.


%

\section{Summary and conclusions}
\label{Summary}


In this work we have performed a two and three dimensional Bayesian study for parameter estimation of the extended $\sigma$-$\omega$ EoS model of neutron stars. The chosen EoS parameters under study are the
compressibility of nuclear matter $K_0$, the symmetry energy $S_0$ and the Landau mass $m_L$, properties of saturation density dense matter. These parameters where varied within an acceptable range of empirical values. It is well known that compressibility of nuclear matter dominates the stiffness of neutron star matter therefore significantly contributes to the determination of the maximum neutron star mass $M_{max}$. On the contrary, the symmetry energy has a greater influence on the determination of the neutron star radius $R$~\cite{Chamel:2019hml,Kubis:2012pha,Klahn:2006ir} rather than on $M_{max}$.
For our analysis we have used as observational inputs for our Bayesian study the multi-messenger astronomy measurements: mass measurements of PSR\,J0740+6620~\cite{Cromartie:2019kug} and PSR\,J0348+0432~\cite{Antoniadis:2013pzd}, tidal deformability data from GW170817~\cite{TheLIGOScientific:2017qsa,Abbott:2018exr}, $M_{max}$ boundaries~\cite{Most:2018hfd}, and combined mass-radius measurements for  PSR J0030+0451~\cite{Miller:2019cac,Riley:2019yda}.

An important element in our Bayesian analysis is that the choice of the corresponding prior distributions for these astrophysical measurements have been taken as uniform. We have observed that pulsar maximum mass measurements favour any stiff EoS models with a high maximum mass value above 2 M$_{\odot}$. On the other hand, gravitational wave data both in the form of neutron star tidal deformabilities as well as the excluded high mass region do not favour the stiffest EoS models in our sample. Thus, there is an existing an interplay between these constraints. 

The Landau mass $m_L$ is the best determined parameter whereas the remaining two have probabilities distributions peak at the lowest values we have considered. Therefore, the values of $K_0$ and $S_0$ found by our analysis have been compared with empirical values found in the laboratory. We find that probability distributions do not exactly peak at their measured value, however they lie within the one sigma confidence region of their posterior distribution. Moreover, their influence on the neutron star properties results to be negligible with respect to $m_L$. Our one and two parameter study corroborates that it is reliable to set $K_0$ and $S_0$ within their empirical values, therefore entrusting the impact on neutron stars to the Landau mass $m_L$ whose values are inversely proportional to $M_{max}$. Moreover, the Bayesian study of~\cite{Margueron:2017lup} considers an inversion of the TOV equations in order to study a set of nuclear parameters derived from an expansion of the EoS functional under three different scenarios for DUrca cooling activation. For the first
set DUrca is activated for masses below 2 M$_{\odot}$, for their intermediate set DUrca holds within 1.8 $< M/M_{\odot}<$ 2, and within 1.8 $< M/M_{\odot}<$ 2 for the last one. They corroborate the hypothesis of a universal contribution from the symmetry energy under the DUrca constraint~\cite{Blaschke:2016lyx}  and find the corresponding results for these three categories: a) $K_0=232.5\pm18.0$ MeV \& $S_0=31.9\pm2.0$ MeV, b) $K_0=231.7\pm18.3$ MeV \& $S_0=31.6\pm1.9$ MeV and, c) $K_0=231.6\pm18.2$ MeV \& $S_0=31.8\pm3.8$ MeV, values that fully fall within our study range.
 
All in all, we find that within the set of most probable neutron star sequences the values of 13 km $<R_{1.4}<$ 13.5 km and $M_{max} \approx 2.2$ M$_{\odot}$  hold for the stiffest EoS models. The latter quantity is in agreement with estimations on the bound for the maximum neutron star mass from rotation studies of the remnant of the merger in GW170817~\cite{Most:2018hfd} as is shown in the excluded region of the $M$-$R$ diagram of Fig.~\ref{MR_constraints}.

\section{Acknowledgments}
D. A-C. acknowledges support from the the Bogoliubov-Infeld program for collaboration between JINR and Polish Institutions as well as from the COST actions CA15213 (THOR) and CA16214 (PHAROS). G. G. B. and P. P. were supported by Hungarian National Research Fund NKFIH (OTKA) grants K120660, K123815, 2019-2.1.11-TET-2019-00050, 2019-2.1.11-TET-2019-00078, COST actions CA15213 (THOR) and CA16214 (PHAROS). A. A. and and H. G. were supported by the RFBR grant No. 18-02-40137. The authors also acknowledge the computational resources of the Wigner GPU Laboratory and of the Laboratory of IT (JINR)~\cite{HybriLIT:2018out}.

\end{document}